\documentclass[prb,aps,twocolumn,amsmath,amssymb,showpacs]{revtex4}

\usepackage{graphicx}
\usepackage{epsfig}
\usepackage[english]{babel}
\usepackage{natbib}
\usepackage{amsfonts}
\usepackage{amssymb}
\usepackage{graphicx}
\usepackage{dcolumn}
\usepackage{amsmath}
\usepackage[ansinew]{inputenc}
\usepackage{psfig}
\usepackage{amssymb}

\begin{document}
\date{\today}
 \title
{Chaos and plasticity in superconductor vortices: a low-dimensional dynamics}
\author
{E. Olive, J.C. Soret}
\affiliation{LEMA, UMR 6157, Universit\'e Fran\c{c}ois Rabelais - CNRS - CEA, Parc de Grandmont 37200 Tours, France}


\begin{abstract}
We present new results of numerical simulations for driven vortex lattices in presence of random disorder at zero temperature. We show that the plastic dynamics of vortices display dissipative chaos. Intermittency "routes to chaos" have been clearly identified below the differential resistance peak. The peak region is characterized by positive Lyapunov exponents characteristic of chaos, and low frequency broad-band noise. Furthermore we find a low fractal dimension of the strange attractor,
which suggests that only a few dynamical variables are sufficient to model the complex plastic dynamics of vortices. 
 \end{abstract}

\maketitle

\section{Introduction} 
\label{section1}
The physics of periodic structures in a random pinning potential has significantly improved these last years. In particular, an elastic approach of the static problem has brought conclusive results in many situations \cite{Larkin,TGPLD1,Carpentier-Kierfeld-Fisher,Klein,Gingras-Ryu-VanOtterlo}.\\ 
When driven over the random medium, the fast moving periodic structure is able to absorb dislocations that eventually appeared at intermediate velocities \cite{Thorel,Moon-Faleski-Olson-Fangohr-Chandran,Kolton}. In their elastic approach, Giamarchi and Le Doussal \cite{TGPLD2} show that a non-linear static disorder persists in the direction transverse to motion even at high velocity. They therefore conclude that the moving lattice is a moving glass and not a moving crystal. As in the static case, in dimension three and weak disorder the relative transverse displacements only grow logarithmically above a given length scale. The moving glass is therefore characterized by a quasi long range order with perfect topological order and rough static channels in which "particles" are flowing. In the case of point disorder, such moving structure is called the {\it moving Bragg glass}. Dislocations between the channels are expected for stronger disorder or in dimension two, creating {\it decoupled channels} and the resulting moving glass is called a {\it moving transverse glass}. It is characterized by a smecticlike order transverse to the direction of motion. However, the question of the stability of the fast moving phase is still controversial \cite{Balents}. In the particular case of correlated disorder created for example by heavy ions irradiation in type II superconductors, the vortex moving structure is a {\it moving Bose glass} \cite{Chauve} characterized by a transverse critical force and a diverging tilt modulus due to the localization effect arising from the columnar pins. This novel feature specific to correlated disorder results in the so-called {\it dynamical transverse Meissner effect} that has been confirmed in numerical simulations \cite{Olive1}.

In the case of strong point disorder or intermediate velocities dislocations are likely to appear and plastic deformations should be considered. In conventional superconductors the vortex experiments show plasticity close to the melting line, {\it i.e.} in the {\it peak effect} region. In this region of the phase diagram, the shape of the current-voltage $I-V$ curve is modified and displays a peak in the differential resistance $dV/dI$ curve \cite{Bhattacharya,Higgins-Paltiel}. Furthermore, the voltage noise measurements increase of several orders of magnitude in this region \cite{Higgins-Paltiel,Marley}. Recent studies explain the peak effect with the important role of surface pinning or surface barriers that lead to non-uniform current flow in the sample \cite{Paltiel-Marchevsky-Pautrat-Simon}.
In numerical simulations, the plasticity is observed when the pinning is not too weak leading to filamentary depinning made of plastic flow around pinned regions \cite{Jensen-Koshelev-GronbechJensen-Ryu-Spencer,Moon-Faleski-Olson-Fangohr-Chandran,Kolton}. In this case the intermediate velocity regime is accompanied by a change of curvature in the velocity-force curve, generating a peak in the derivative curve. This is similar to the peak of the differential resistance curve obtained in the {\it peak effect} region of type II superconductors. In such plastic phases topological defects proliferate in the vortex lattice, and a complex and apparently very disordered dynamics of vortices is growing in size. It is therefore clear that these regimes cannot be described by an elastic approach. The theoretical understanding of plastic depinning and plastic flows remains an open problem. Different coarse-grain models (where the Larkin domains are the degrees of freedom) are developed to describe plastic deformations: see Ref.\onlinecite{Saunders} and references therein. In particular, Marchetti {\it et al.} \cite{Marchetti} show the existence of a tricritical point which separates continuous depinning transitions at weak disorder from discontinuous depinning transitions with hysteresis at strong disorder. These mean field results agree with the vortex simulations of Ref.\onlinecite{Olson2}. However, numerical studies of a phase slip model \cite{Nogawa} suggest the absence of hysteresis at the thermodynamic limit. In another numerical phase slip model, the plastic depinning for strong disorder appears to be a continuous transition \cite{Kawaguchi}.

The theoritical approach of plastic flows is difficult due to their intrinsic complexity. 
However, we recently showed \cite{Olive2} that a natural approach in terms of dissipative chaos gives some new understandings of the vortex plasticity. In particular, a low dimensional dynamics has been evidenced for $N_v=30$ vortices. In the present paper we show the pertinence of the chaotic approach for larger systems, but more importantly we show that the low fractal dimension of the chaotic attractor remains unchanged despite a much higher dimensional phase space.\\
The outline of this paper is as follows. In section \ref{section2} we describe our numerical model for line vortices in type II superconductors. In section \ref{section3} we show in details the transition to chaos followed by the vortices in the plastic phase. Section \ref{section4} characterizes the chaotic phase itself which is evidenced in a wide range of the driving force around the differential resistance peak. Positive Lyapunov exponents, broad-band noise at low frequency and fractal dimension of the strange attractor are analysed in details. The crucial conclusion of a low dimensional dynamics is explained, and the influence of the number of degrees of freedom and dissipation are discussed in section \ref{section5}.

\section{Numerical model} 
\label{section2}

We consider $N_v$ Abrikosov vortices driven over a random pinning background in the $(x,y)$ plane.
At $T=0$ the overdamped equation of motion of a vortex $i$ in position $\bold r_i$ reads 
   \begin{eqnarray}
\eta {{d{\bf r_i}}\over{dt}}=-{\sum_{j \neq i}} \boldsymbol\nabla_i U^{vv}(r_{ij})-{\sum_{p}}\boldsymbol\nabla_i U^{vp}(r_{ip})+{\bf F}^L
\label{eq1}  
 \end{eqnarray}
where $r_{ij}$ is the distance between vortices $i$ and $j$, $r_{ip}$
 is the distance between the vortex $i$ and the pinning site located at ${\bf r_p}$, and $\boldsymbol\nabla_i$ is the 2D gradient operator acting in the 
$(x,y)$ plane. The viscosity coefficient is $\eta$, ${\bf F}^L=F^L{\bf \hat x}$ is the Lorentz driving force due to an applied current.
The vortex-vortex pairwise repulsive interaction is given by a modified Bessel function 
\begin{eqnarray}\nonumber  
U^{vv}(r_{ij})=2\epsilon_0A_v K_0(r_{ij}/\lambda_L),
\end{eqnarray}
 and the attractive pinning potential is given by 
\begin{eqnarray}\nonumber  
U^{vp}( r_{ip})=-\alpha_p e^{-(r_{ip}/R_p)^2}.
\end{eqnarray}
 In these expressions $A_v$ and $\alpha_p$ are tunable parameters, $\lambda_L$ is the magnetic penetration depth, and $\epsilon_0=(\phi_0/4\pi \lambda_L)^2$ is an energy per unit length. 
We consider periodic boundary conditions of $(L_x, L_y)$ sizes in the $(x,y)$ plane. All details about our 
method for computing the Bessel potential with periodic conditions can be found in Ref.\onlinecite{Olive}. A Runge-Kutta algorithm for molecular dynamics simulation is used for $N_v=30,\ 270$ and $1080$ 
vortices in a rectangular basic cell $(L_x,L_y)=(5, 6\sqrt 3/2)n\lambda_L$, where $n=1,\ 3$ and $6$ respectively. 
We choose the same density of pinning centers as the vortex density. We consider the London limit $\kappa =\lambda_L /\xi =90$, where $\xi$ is the superconducting coherence length. $\xi$ appears in the inner cutoff which removes the logarithmic divergence of $K_0(r/\lambda_L)$ at $r\rightarrow 0$ (see Ref. \onlinecite{Olive}). The average vortex distance $a_0$ is set to $a_0=\lambda_L$, and  $R_p=0.22\ \lambda_L$, $\eta=1$, $A_v=2.83\times 10^{-3}\lambda_L$. We computed different pinning strengths $\alpha_p/A_v\sim 0.05$, $\alpha_p/A_v\sim 0.35$ and $\alpha_p/A_v\sim 1.05$,
corresponding to a maximum pinning force of respectively $F_{max}^{vp}\sim 0.2F_0$ , $F_{max}^{vp}\sim 1.4F_0$ and $F_{max}^{vp}\sim 4F_0$,  
where $F_0=2\epsilon _0A_v/\lambda_L$ is a force defined by the Bessel interaction.
The driving force applied along a principal vortex lattice direction $x$ is slowly varied from $0$ up to a value far from the critical Lorentz force along $x$.\\
The successive regimes we observe are pinned configurations where all vortices have zero velocity, followed by plastic channels flowing through pinned regions. The successive depinnings of the pinned regions result in a complex plastic phase of interconnected flowing channels. Finally coupled or decoupled channels mostly aligned with the driving force appear in the high driving phase. \\

\section{Intermittency "route to chaos"} 
\label{section3}

\begin{figure}
\includegraphics[width=0.86\linewidth]{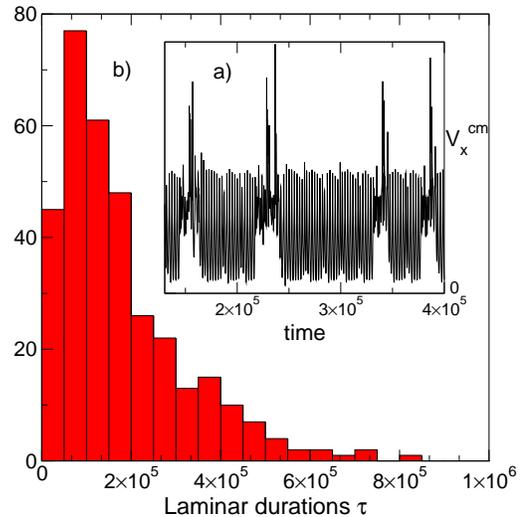}
\caption{ \label{fig1} (Color online) Intermittency route to chaos characteristics observed for $N_v=30$ vortices and for a pinning strength $\alpha_p/A_v\sim 0.35$ : a) part of the time evolution of the longitudinal velocity $V_x^{cm}(t)$ obtained for $F^L=1.116\times 10^{-3}$. One sees laminar ({\it i.e.} periodic) phases interupted by chaotic bursts of large erratic velocity fluctuations. b) distribution of the laminar phase durations of $V_x^{cm}(t)$. Both plots indicate type II intermittency (see text).
}
\end{figure}

We first begin with the transition to chaos (often called {\it routes to chaos} in dynamical system theory) that we observe for vortices in a very short applied force range below the differential resistance peak. In particular we show the results we obtain for $N_v=30$ vortices and for the pinning strength $\alpha_p/A_v\sim 0.35$. In Fig.\ \ref{fig1}.a we show the typical longitudinal velocity of the vortex center of mass $V_x^{cm}(t)$ that we measure in time for a given Lorentz force $F^L=1.116\times 10^{-3}\sim 0.3943F_0$. 
$V_x^{cm}(t)$ shows time intervals where the motion is almost periodic (laminar phases) and interrupted by chaotic bursts displaying large erratic velocity fluctuations. The almost periodic motion corresponds to a plastic channel flowing through pinned regions, and the instability gives way to a chaotic burst where all vortices are moving erratically.
Later the system goes back to the almost periodic regime with pinned vortices and moving vortices that synchronize temporarily their motion. Again the instability gives way to another chaotic burst, and so on. Therefore, such signal $V_x^{cm}(t)$ shows intermittency which may be one of the three main known scenarios to drive a dissipative system from periodicity to chaos \cite{Pommeau,Berge}. In the case of intermittency, the dynamical system has a stable limit cycle below the intermittency threshold $F_t$, whereas for $F\rightarrow F_t^+$ the dynamical regime is intermittent with apparently periodic oscillations interrupted by large fluctuations. The amplitude and duration of these large fluctuations are almost the same from one fluctuation to the other, and they depend little on the force. When $F\rightarrow F_t^+$ these fluctuations become increasingly rare and disappear completely 
below the threshold. Therefore the mean frequency of the chaotic bursts goes to zero at the transition but not the amplitude neither their duration. The intermittency "route to chaos" has several characteristics and may be mainly classified in three types (I, II and III) depending on the way the limit cycle loses stability. In general a trajectory in phase space is linearly stable if all the eigenvalues (called the Floquet multipliers) of the Floquet matrix have magnitude smaller than $1$, and loses stability if one of them leaves the unit circle in the complex plane \cite{Berge}. A crossing by the $(+1)$ value on the real axis generates a saddle-node bifurcation and gives type I intermittency. The crossing by $(-1)$ on the real axis generates a a subcritical subharmonic bifurcation and gives type III intermittency, whereas type II intermittency is observed above a subcritical Hopf bifurcation and corresponds to a crossing of the Floquet multiplier by two conjugate complex values. 
To determine the type of intermittency corresponding to Fig.\ 1a, we first measure for a given value of the applied force the distribution of the laminar ({\it i.e.} periodic) phase durations. Fig.\ 1b shows such distribution obtained for $V_x^{cm}(t)$ displayed in Fig.\ 1a. Such distribution is qualitatively very different from the one obtained for weaker pinning $\alpha_p/A_v\sim 0.05$
 corresponding to $F_{max}^{vp}\sim 0.2F_0$ 
shown in Ref.\onlinecite{Olive2}, and characteristic of type I intermittency. The very different shape of the distribution we observe here for $\alpha_p/A_v\sim 0.35$, in particular the long tail for large durations, cannot be attributed to type I intermittency but is expected for type II or type III intermittency. A possible way to discriminate between these two possibilities is to enlarge the signal $V_x^{cm}(t)$ at the end of the almost periodic intervals, {\it i.e.} just before the chaotic bursts. Since type III intermittency is associated to a period doubling bifurcation (crossing by the $(-1)$ value of the Floquet multiplier), the chaotic burst is expected to appear just after the increasing of a $1/2$ subharmonic oscillation at the end of the laminar intervals. Such behavior is clearly not observed in the signal $V_x^{cm}(t)$ displayed in Fig.\ 1a., and is confirmed by a power spectrum analysis. 
We therefore conclude that type II intermittency is the "route to chaos" followed by $N_v=30$ vortices in rather strong pinning $\alpha_p/A_v\sim 0.35$ \cite{note}. Since type I intermittency was observed for weaker pinning $\alpha_p/A_v\sim 0.05$
\cite{Olive2}, we therefore show that the pinning strength may change the nature of the bifurcations brought into play.\\

\section{Chaos} 
\label{section4}

We now turn to the chaotic phase itself wich occurs in a large force range above the very narrow intermittency range. For the study of chaos we now concentrate on a larger system size of $N_v=270$ vortices flowing over a random medium made of $N_p=270$ pinning centers. We present results for a pinning strength of $\alpha_p/A_v\sim 1.05$. 

\subsection{Differential resistance}
\label{section4-ssection1}

\begin{figure}
\includegraphics[width=0.86\linewidth]{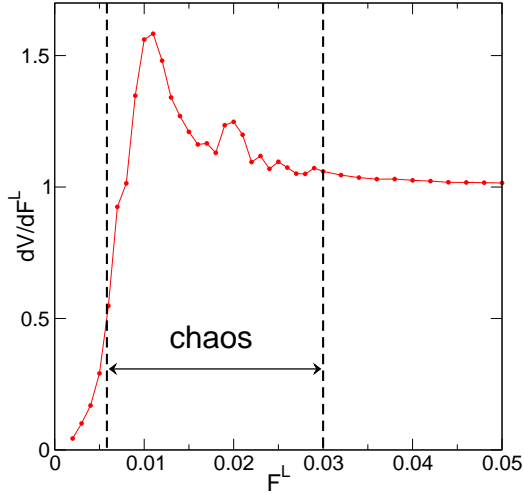}
\caption{ \label{fig2} (Color online) Differential resistance $R_d=dV_x^{cm}/dF^L$ obtained for $N_v=270$ vortices and for a pinning strength of $\alpha_p/A_v\sim 1.05$\ .
The main peak close to $F^L=0.01$ indicates the {\it S} shape of the velocity-force curve, and the so-called {\it fingerprint} phenomenon is observed between $F^L=0.02$ and $F^L=0.03$. As indicated by doted lines, chaos is measured between $F^L=0.006$ and $F^L=0.03$, as shown in Fig.\ \ref{fig3}.}
\end{figure}

Figure \ref{fig2} displays the differential resistance curve $R_d=dV_x^{cm}/dF^L$. The shape is very similar to the differential resistance curve obtained in the {\it peak effect} region of type II superconductors \cite{Bhattacharya,Higgins-Paltiel}. It shows a peak corresponding to the {\it S} shape of the velocity-force curve. In numerical simulations, this peak indicates plasticity in the vortex flowing. Above the main peak in Fig.\ \ref{fig2} appears the so-called {\it fingerprint} phenomenon already observed in experiments \cite{Bhattacharya,Higgins-Paltiel}. This anomaly corresponds to the depinning of the last pinned vortices. Our study shows that for various system sizes and various pinning strengths, the plasticity of the filamentary vortex flow indicates chaotic dynamics. Chaos begins close to the bottom of the differential resistance peak where the intermittency route to chaos is observed, and ends just above the {\it fingerprint} phenomenon.

\subsection{Lyapunov exponents}
\label{section4-ssection2}

\begin{figure}
\includegraphics[width=0.86\linewidth]{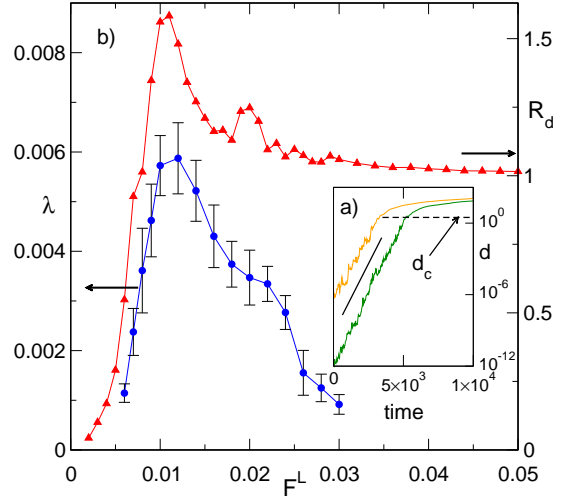}
\caption{ \label{fig3} (Color online) Signature of chaos in the plastic phase for $N_v=270$ vortices and for a pinning strength of $\alpha_p/A_v\sim 1.05$
: a) semi-log plot of the time evolution of the distance $d(t)$ between two initial neighbouring trajectories in phase space for $F^L=0.012$. Two different initial distances between neighbouring trajectories are shown, $10^{-12}$ (green) and $10^{-6}$ (orange). The positive slope $\lambda$ (see solid line as guide for the eye) 
indicates the exponential divergence of $d(t)$ representative of chaos. The same slope observed up to the same length scale $d_c$ above which the growth of $d$ saturates, indicate that both $\lambda$ and $d_c$ are characteristic quantities of the intrinsic chaotic dynamics.  
b) Evolution of the maximal Lyapunov exponent $\lambda$ (blue circles) with the Lorentz force, plotted together with the differential resistance curve $R_d=dV_x^{cm}/dF^L$ (red triangles). Each Lyapunov exponent is the average over 20 couples of initial conditions and the error bars are the standard deviation. }
\end{figure}

An unambiguous signature of chaos is given by at least one positive Lyapunov exponent illustrating the {\it sensitive dependence on initial conditions} (SDIC), which is a property of chaotic attractors only. Starting from two neighbouring initial conditions on the chaotic attractor, the two corresponding trajectories in phase space will diverge exponentially fast one from another while still staying on the chaotic attractor. The Lyapunov exponents are the inverse of the characteristic times of this exponential divergence. Therefore, to compute the maximal Lyapunov exponent $\lambda$ one has to compute the distance $d(t)$ between two initial neighbouring trajectories on the chaotic attractor. Since we integrate $N_v$ first order differential equations of motion (Eq.\ \ref{eq1}), the phase space is defined by the $2N_v$ vortex coordinates 
and the distance $d$ is defined by
\begin{eqnarray}\nonumber  
d^2(t)={\sum_{i=1}^{N_v}}\left[\left(X_i(t)-\tilde X_i(t)\right)^2+\left(Y_i(t)-\tilde Y_i(t)\right)^2\right]
\end{eqnarray}
 where $X_i(t)=x_i(t)-x_{cm}(t)$, $Y_i(t)=y_i(t)-y_{cm}(t)$, $\tilde X_i(t)=\tilde x_i(t)-\tilde x_{cm}(t)$, $\tilde Y_i(t)=\tilde y_i(t)-\tilde y_{cm}(t)$. In these expressions, $(x_i,\ y_i)$ and $(\tilde x_i,\ \tilde y_i)$ are the vortex $i$ coordinates, 
and $(x_{cm},\ y_{cm})$ and $(\tilde x_{cm},\ \tilde y_{cm})$ are the respective coordinates of the center of mass. The tilde notation ($\tilde x,\tilde y$) refers to the second trajectory generated by the neighbouring initial condition. 
Figure\ \ref{fig3}.a displays two examples of the time evolution of $d$ that we typically find in the region of the differential resistance peak. 
It clearly shows an exponential divergence $d\sim e^{\lambda t}$ of the distance between two trajectories starting with an initial distance of $10^{-12}$ between them (see the green curve). 
The slope therefore defines the positive maximal Lyapunov exponent $\lambda$ characteristic of chaotic dynamics. 
When the distance $d$ becomes comparable to the size of the chaotic attractor, a saturation effect in the growth of $d(t)$ naturally appears. Furthermore Fig\ \ref{fig3}.a shows that, for the same driving force, changing the initial distance between the two trajectories (see the orange curve) doesn’t change the maximal Lyapunov exponent $\lambda$ (whithin the error bars), but also doesn't change the length scale $d_c$ below which the exponential divergence is observed, so that both $\lambda$ and $d_c$ appear as characteristic quantities of the intrinsic chaotic dynamics of the vortex system. 
Fig.\ \ref{fig3}.b displays the evolution of $\lambda$ that we measure in the chaotic region. In the same graph we plot the differential resistance curve $R_d=dV_x^{cm}/dF^L$. The shape of the Lyapunov curve closely follows the differential resistance curve. In particular a peak appears in the Lyapunov curve simultaneously with the differential resistance peak, and a shoulder at $F^L\sim 0.2$ indicates the {\it fingerprint} phenomenon shown in Fig.\ \ref{fig2}.
The maximal Lyapunov exponent is in some sense a measure of the degree of chaos since the higher Lyapunov exponent $\lambda$, the faster the divergence of two chaotic trajectories. Therefore Fig.\ \ref{fig3}.b shows that chaos is fully developed at the differential resistance peak. Furthermore, the shoulder in the Lyapunov curve close to $F^L=0.020$ indicates that the decreasing of chaos above the differential resistance peak is slowing simultaneously with the {\it fingerprint} phenomenon  which confirms the idea of the depinning of the last pinned vortices.\\

\subsection{Low frequency noise}
\label{section4-ssection3}

\begin{figure}
\includegraphics[width=0.43\linewidth]{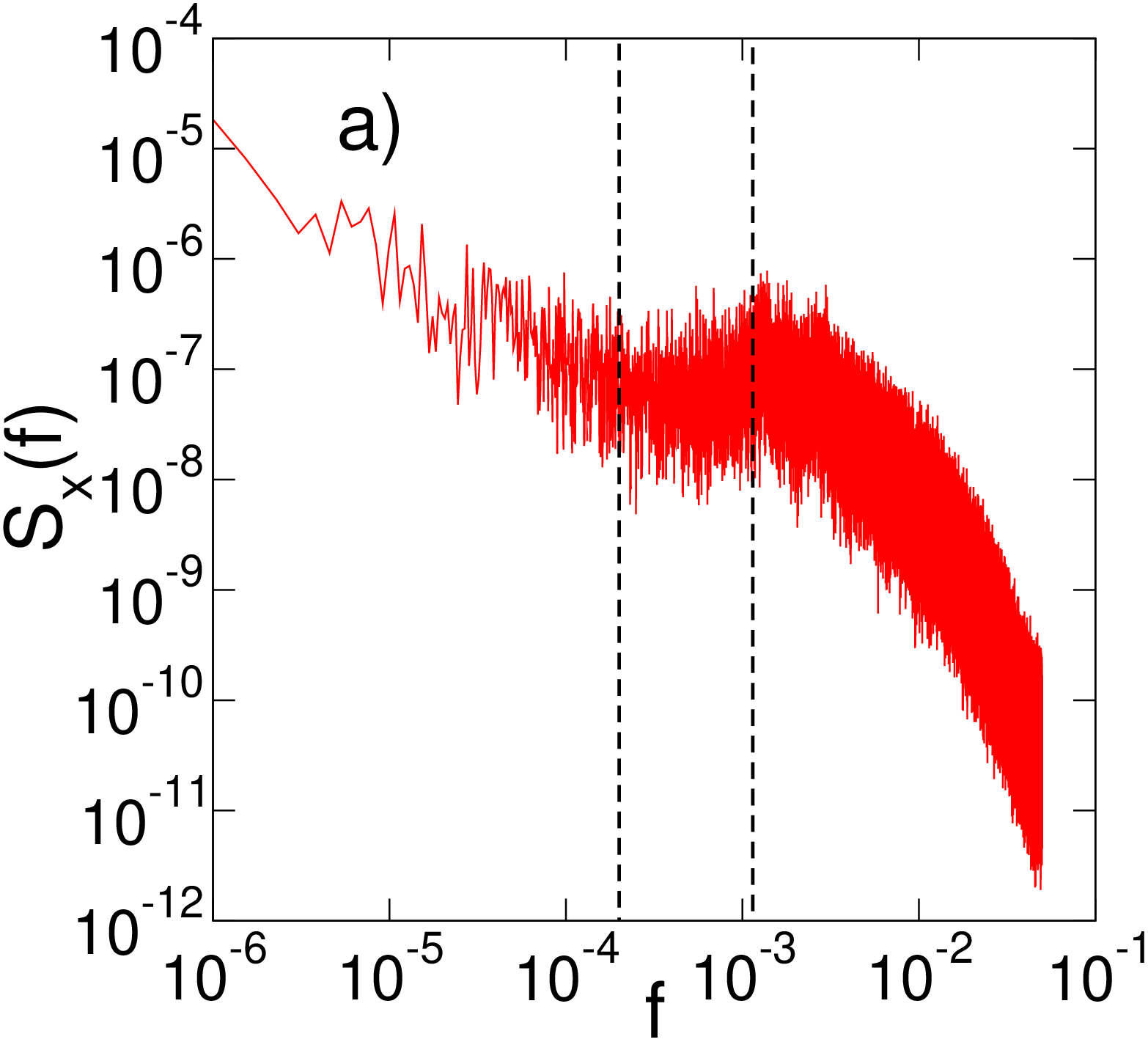}
\includegraphics[width=0.43\linewidth]{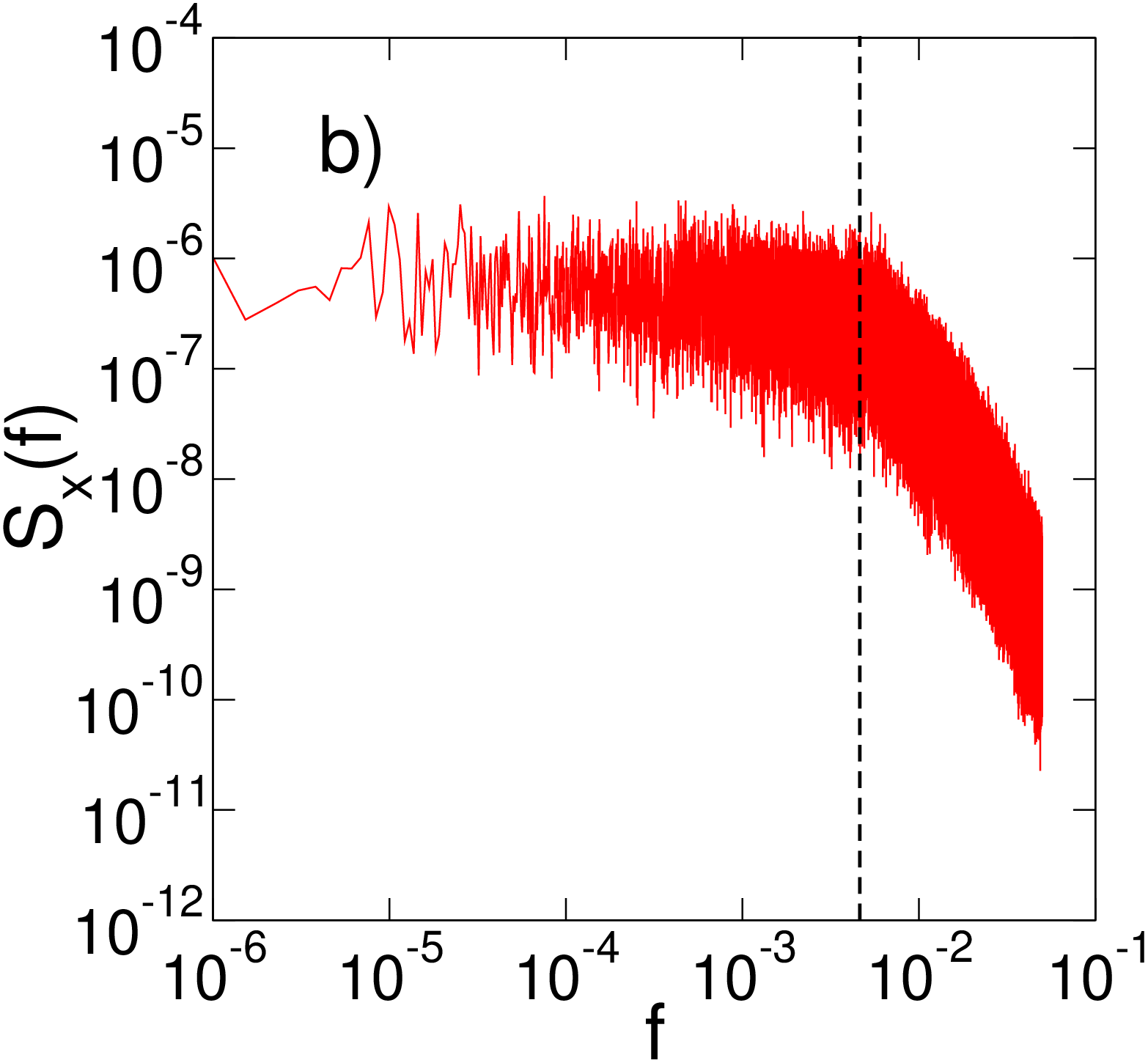}
\caption{ \label{fig4} (Color online) Power spectrum $S_x(f)$ of the longitudinal noise for $N_v=270$ vortices and for a pinning strength of $\alpha_p/A_v\sim 1.05$:
 a) $F^L=0.006$, below the differential resistance peak. The lower dotted line corresponds to $1/\tau_{diff}$ while the higher corresponds to the maximal Lyapunov exponent $\lambda$. b) $F^L=0.009$ corresponding to the peak of the differential resistance. The dotted line corresponds to the maximal Lyapunov exponent $\lambda$. In this case, $\tau_{diff}$ is larger than the experimental time window so that $1/\tau_{diff}$ does not appear in the power spectrum. The chaotic noises $N_x$ and $N_y$ shown in Fig.\ \ref{fig5} are averaged in the range $1/\tau_{diff}<f<\lambda$. For $F^L=0.006$, note the colored noise signature $S_x(f)\sim f^{-\gamma}$ of the diffusion process in the power spectrum for $f<1/\tau_{diff}$.}
\end{figure}

\begin{figure}
\includegraphics[width=0.86\linewidth]{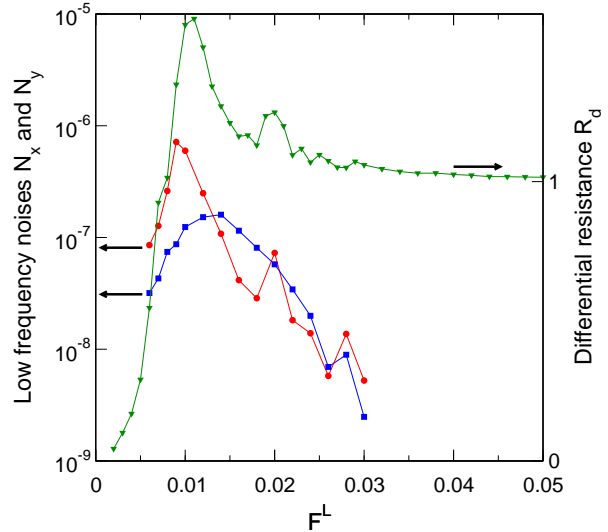}
\caption{ \label{fig5} (Color online) $N_v=270$ vortices and for a pinning strength of $\alpha_p/A_v\sim 1.05$
: differential resistance (green triangles), and longitudinal $N_x$ (red circles) and transverse $N_y$ (blue squares) low frequency noise averaged in the range $1/\tau_{diff}<f<\lambda$.}
\end{figure}

Many experiments in type II superconductors have measured large excess noise, in particular in the {\it peak effect} region \cite{Higgins-Paltiel,Marley} or close to the first order transition where the vortex lattice melts in the vortex liquid phase \cite{Togawa}, and very few show transverse noise measurements \cite{Maeda-Scola}. However the origin of the excess noise remains an open question since several mechanisms are proposed. In numerical simulations, the differential resistance peak and excess noise originate from plastic flow of vortices. We analyse plasticity in terms of chaos and in particular we argue that the low frequency broad-band noise measured in the differential resistance peak region is a signature of the SDIC property which is characteristic of chaos since it implies the loss of memory for the chaotic system.\\
We compute the power spectrum $S_{\alpha}(f)$ of the velocity of the center of mass $V_{\alpha}^{cm}(t)$ defined by
\begin{eqnarray}\nonumber  
S_{\alpha}(f)={{1}\over{t_2-t_1}} \left| \int_{t_1}^{t_2}dt\ V_{\alpha}^{cm}(t)\ exp(i2\pi ft )\right| ^2
\end{eqnarray}
where $\alpha=x$ or $y$. Figure \ref{fig4} shows the typical power spectra we obtain close to the differential resistance peak. 
The broad-band noise at low frequency is obvious and shows the impossibility of long term prediction. However, the {\it washboard} frequency $f=V_x^{cm}/a_0$ appears in the power spectra immediately above the peak of the differential resistance (not shown) therefore showing the beginning of temporal order of the lattice although still chaotic. 
We shall now compute the broad-band low frequency noise characteristic of chaos. Two time scales appear in the region of the differential resistance peak. $1/\lambda$, where $\lambda$ is the Lyapunov exponent, is the characteristic time above which chaos appears and $\tau_{diff}$ is the characteristic time above which diffusive motions are measured. We compute $w_x(t)={1\over N_v}{\sum_{i=1}^{N_v}}\left[X_i(t)-X_i(0)\right]^2$ and 
$w_y(t)={1\over N_v}{\sum_{i=1}^{N_v}}\left[Y_i(t)-Y_i(0)\right]^2$, with $X_i(t)=x_i(t)-x_{cm}(t)$ and $Y_i(t)=y_i(t)-y_{cm}(t)$. As already found by Kolton {\it et al.} \cite{Kolton} we find $w_x(t)\sim t^{\xi_x}$ and $w_y(t)\sim t^{\xi_y}$ with exponents indicating normal and anomalous diffusion. We observe these diffusive motions for time scales larger than $\tau_{diff}$, which varies with the driving force: a maximum of $\tau_{diff}$ is found and coincides with the differential resistance peak. Such diffusive motions have a clear signature in the power spectra $S_{\alpha}(f)$. Indeed, for very low frequencies corresponding to time scales larger than $\tau_{diff}$ a colored noise $S_{\alpha}(f)\sim f^{-\gamma}$ appears in the power spectrum (see figure \ref{fig4}a). 
Therefore, to measure the degree of chaos we compute the average noises $N_x$ and $N_y$ over the low frequency range $1/\tau_{diff}<f<\lambda$, {\it i.e.}
\begin{eqnarray}\nonumber  
N_{\alpha}={{1}\over{\lambda-{1/\tau_{diff}}}} \int_{1/\tau_{diff}}^{\lambda}\ S_{\alpha}(f)\ df
\end{eqnarray}
where $\alpha=x$ or $y$, and the Lyapunov exponent $\lambda$ have been evidenced in Fig.\ \ref{fig3}. 
Concomitantly with the differential resistance, Fig.\ \ref{fig5} shows the longitudinal $N_x$ and transverse $N_y$ low frequency noises averaged in this way in the chaotic region. 
Again the shape of the noise curves closely follows the differential resistance curve. The maximun of the longitudinal noise coincides with the peak of the differential resistance which confirms that chaos is fully developed at the peak. Furthermore, anomalies in the low frequency noises appear simultaneously with the {\it fingerprint} phenomenon. Finally note that the maximal longitudinal noise level is almost three orders of magnitude higher than the level measured at the end of the chaos region, {\it i.e.} just above the {\it fingerprint} phenomenon.

\subsection{Fractal dimension of the strange attractor, and low-dimensional dynamics}
\label{section4-ssection4}

\begin{figure}
\includegraphics[width=0.86\linewidth]{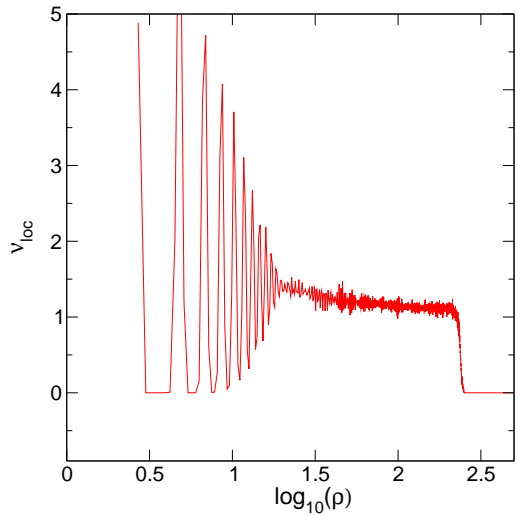}
\caption{ \label{fig6} (Color online) Local slope $\nu_{loc}$ with respect to $log_{10}(\rho)$ for $F^L=0.008$ in the case of $N_v=270$ vortices and a pinning strength of $\alpha_p/A_v\sim 1.05$
. A constant value of $\nu_{loc}$ appears showing the validity of the power law $C(\rho) \sim \rho ^\nu$. The value $\nu\sim 1.12$ for the correlation dimension is measured for $F^L=0.008$ in the power law regime. Below and above this regime, are two regimes that clearly deviate from the power law (see the text for details).}
\end{figure}

In sections  \ref{section4-ssection2} and \ref{section4-ssection3} we characterized the absence of temporal correlation in the chaotic regime due to the SDIC fundamental property. We shall now characterize the spatial correlations within the chaotic regime by computing the dimension of the chaotic attractor. The peculiarity of a chaotic attractor comes from its two properties that seem hard to reconcile. Trajectories of phase space converge and remain confined to a bounded region (attraction), although neighbouring trajectories on the attractor separate exponentially fast (SDIC). It involves a double mechanism:  {\it stretching} necessary for the SDIC property, followed by {\it folding} required for confined trajectories to a bounded region. Repeated applications of stretching and folding generate a fractal attractor whose peculiar properties justify the name {\it strange attractor}. There exist several definitions of a fractal dimension which are not equivalent and lead to different numerical values for a given object.
The Hausdorff-Besicovitch dimension is probably the most famous but rather unworkable in practice. A more efficient method is to compute the correlation dimension proposed by Grassberger and Procaccia \cite{Grassberger}. Consider a set of very many points $\{X_i(t),Y_i(t)\}$ (see section \ref{section4-ssection2}) on the strange attractor generated by letting the system evolve a long time. The correlation dimension $\nu$ is the exponent of the power law $C(\rho) \sim \rho ^\nu$, where 
\begin{eqnarray}\nonumber  
C(\rho)={lim}_{m\rightarrow \infty}{1\over m^2}{\sum_{k,l=1}^m}H(\rho-\rho_{kl})
\end{eqnarray}
measures the number of couples of points $(k,l)$ on the chaotic attractor whose distance $\rho_{kl}$ is less than $\rho$. $H(z)$ is the Heaviside function. To estimate the correlation dimension $\nu$, we determine the local slope on a log-log plot of $C(\rho)$ defined by 
\begin{eqnarray}\nonumber  
\nu_{loc}={d(log_{10}C(\rho))\over d(log_{10}\rho))}\ . 
\end{eqnarray}
$\nu$ is deduced from the intervals where $\nu_{loc}$ is constant, {\it i.e.} where a true power law is measured. Figure \ref{fig6} displays $\nu_{loc}$ with respect to $log_{10}\rho$ that we obtain for $F^L=0.008$.
It clearly shows a constant value of $\nu_{loc}\sim 1.12$ over a limited range of $\rho$. It shows the validity of the power law behavior $C(\rho) \sim \rho ^\nu$ in this range. The correlation dimension for $F^L=0.008$ is therefore $\nu\sim 1.12$. Above and below the limited range where the power law behavior is observed, two regimes exist that are well known for such correlation dimension computations. Because of the finite number of points $\{X_i(t),Y_i(t)\}$ defining the strange attractor in our simulations, the number of pairs $(k,l)$ whose distance is less than $\rho$ is small for small values of $\rho$, giving therefore a poor statistics. And above the limited range where $\nu_{loc}$ is constant, $\rho$ becomes comparable to the size of the strange attractor, and increasing $\rho$ will not add new $(k,l)$ couples. Therefore $C(\rho)\rightarrow 1$ and such saturation leads to $\nu_{loc}\rightarrow 0$.\\
The computation of the correlation dimension in the whole chaotic region, {\it i.e.} in the region of the differential resistance peak, shows the important result that 
\begin{eqnarray}\nonumber  
1<\nu<1.5\ . 
\end{eqnarray}
This result indicates that in our simulations where the time is discrete the chaotic trajectories in phase space remain confined in a subspace of dimension less than two. Since the discrete time system can be considered as a Poincaré section (dimension $D-1$) of the continuous time system (dimension $D$), we conclude that the continuous time system of differential equations (Eq.\ \ref{eq1}) has chaotic trajectories confined in a subspace of dimension less than three. This is a crucial result since it indicates that the complex plastic vortex dynamics of very many degrees of freedom is low dimensional, and can be modeled by only three dynamical variables. \\
Note that the fractal dimension of the chaotic attractor that we find is strikingly similar to the fractal dimension of the vortex channel network in the 2D real space found in Ref.\ \onlinecite{Olson98} for different pinning strengths. Such coincidence is not clear since the link between chaotic trajectories in phase space and in real space is inaccessible. 

\section{Discussion} 
\label{section5}

The same result $1<\nu<1.5$ was obtained for much less degrees of freedom ($60$ instead of $540$ considered here) in our previous paper \cite{Olive2}. We also checked that for the much larger system $N_v=1080$, {\it i.e.} $2160$ degrees of freedom, the chaotic trajectories remain bounded in a low dimensional fractal attractor with $1<\nu<1.5$. Therefore the important result of a low dimensional dynamics of the continuous time system of differential equations Eq.\ \ref{eq1} on a chaotic attractor of dimension less than three remains valid whatever the number of degrees of freedom $N_{df}$ is.\\
Since the system size does not seem to play any role in the low dimension of the strange attractor, we investigate the role of dissipation which shrinks the volumes in phase space and might influence the dimension of the attractor. Indeed, the dissipation of a dynamical system is given by the divergence of the flow. Since Eq.\ \ref{eq1} is a first order differential equation of the form $\bf {\dot r} = \bf {F(r)}$ the divergence ${\boldsymbol{\nabla}}.{\bf F}$ of the flow is not controlled by the viscosity coefficient $\eta$ but by the vortex-pin and vortex-vortex force component derivatives 
$\sum_{i}\left({\partial{F_i}}/{\partial{x_i}}+{\partial{F_i}}/{\partial{y_i}}\right)$.
Keeping the same parameters $\xi$, $\lambda_L$ and $A_v$ for the line vortices, we can therefore modify the dissipation of the flow by modifying the pinning parameters, {\it e.g.} the pinning strength $\alpha_p$. We therefore computed the fractal dimension of the chaotic attractor for $\alpha_p/A_v\sim 0.05$, {\it i.e.} for a pinning strength 20 times less. In this case the pinning is low since while increasing the driving force for $N_v=30$ vortices we observe a clear transition between chaos (plasticity) and the coupled channels regime that might be called the moving Bragg glass. Our computations show that in the chaotic phase, the same result $1<\nu<1.5$ appears. Furthermore, for the same low dissipative flow where $\alpha_p/A_v\sim 0.05$, and for a much larger system size $N_v=1080$ vortices, the same result $1<\nu<1.5$ holds. These computations therefore clearly show that the low fractal dimension of the chaotic attractor that we measure remains unchanged for larger system sizes $30<N_v<1080$ or/and for lower flow dissipation. 

A natural extension of our present study is to analyse the chaotic dynamics of driven vortices in 3D. Previous 3D studies \cite{Olson00-Kolton00} obtained different dynamical regimes depending on the coupling strength between planes and the driving force. In particular, it would be very interesting to compute the fractal dimension of the chaotic attractor in the 3D plastic and 3D smectic phases.

\section{Conclusions} 
\label{section6}

In this paper we investigated the properties of plastic flows of superconductor vortices driven over a random medium. We interpreted the plastic dynamics in terms of dissipative chaos. The link with the differential resistance peak is established. Chaos begins at the beginning of the peak and ends after the "fingerprint" phenomena that occurs above the peak. We first identified with details the transition to chaos ("route to chaos"). The scenario we have measured in our numerical simulation is the intermittency. Type I and type II intermittency occur depending on the pinning strength. We then studied in details the chaotic dynamics that occurs in the force range of the differential resistance peak. Positive Lyapunov exponents are measured which is a non-ambiguous signature of chaos. A low frequency broad-band noise has also been evidenced in the chaotic region. Both Lyapunov exponents and low frequency broad-band noise closely follow the evolution of the differential resistance curve, in particular the existence of a peak. Finally, we measured the fractal dimension of the chaotic attractor. We showed that the discrete time system in our simulation is confined to a subspace of dimension less than $2$. We therefore conclude that the continuous time system of differential equations Eq.\ \ref{eq1} is confined to a subspace of dimension less than $2+1=3$. It suggests that the plastic dynamics of vortices may be described by a model with only three dynamical variables. We showed that this important result still holds for systems as large as $2160$ degrees of freedom and also for systems with low dissipation. Our results give new understandings of the plastic flow phase and open new perspectives for further theoritical studies of plasticity.


\end{document}